\documentclass[a4paper, oneside, twocolumn, notitlepage, 10pt]{extarticle_ecoc}
\usepackage{ecoc}

\usepackage{tikz}
\usepackage{pgfplots}
\usepackage{pgfplotstable}
\usepackage[nolist]{acronym}
\usepackage{orcidlink}


\begin{document}
\selectlanguage{english}    


\title{Bridging the 6G Gap: Scaling Sustainable ROADM-Based IP-over-WDM via DSCM-Enabled Point-to-Multipoint Designs}%


\author{
    Matin Rafiei Forooshani\textsuperscript{(1)}, Farhad Arpanaei\orcidlink{0000-0003-1061-0614}\textsuperscript{(2*)}, Hamzeh Beyranvand\textsuperscript{(1)}, Mahdi Ranjbar Zefreh\textsuperscript{(3)}, \\ 
    Juan Pedro Fernández-Palacios\textsuperscript{(4)}, Alfonso Sánchez-Macián\textsuperscript{(2)}, Jos\'{e} Alberto Hern\'{a}ndez\textsuperscript{(2)}, \\ and David Larrabeiti\textsuperscript{(2)}
}

\maketitle                  

\begin{strip}
    \begin{author_descr}
    
        \textsuperscript{(1)} Dept of Electrical Engineering, Amirkabir University of Technology (Tehran Polytechnic), Tehran, Iran.
        \textsuperscript{(2)} Dept of Telematic Engineering, Universidad Carlos III de Madrid, 28911 Leganes, Madrid, Spain.\\ 
        \textsuperscript{(3)} CISCO Systems S.R.L., Vimercate (MB), Italy.\\
        \textsuperscript{(4)} Telefónica Research and Development, Madrid, Spain.\\
        \textcolor{blue}{\textsuperscript{(*)} \uline{farhad.arpanaei@uc3m.es}},
        
    \end{author_descr}
\end{strip}

\renewcommand\footnotemark{}
\renewcommand\footnoterule{}


\begin{strip}
    \begin{ecoc_abstract}
  This study compares transponder-based, Point-to-Point, and DSCM-based Point-to-Multipoint (PtMP) access-metro architectures. Findings demonstrate that PtMP IPoWDM significantly optimizes efficiency across diverse geotypes, slashing CAPEX by 92.0\% and power by 99.2\% compared to the traditional benchmark over a ten-year horizon. ©2026 The Author(s) 
    \end{ecoc_abstract}
\end{strip}


\section{Introduction}
\vspace{-0.1cm}
The transition toward the 6G era is reshaping optical transport, imposing stringent requirements on scalability and energy efficiency in hierarchical metro-aggregation networks \cite{ArpanaeiECOC2023}. 6G traffic follows a hierarchical multi-tier aggregation flow from cell sites through central offices to the core \cite{NapoliECOC25}. Traditional transponder-based architectures rely on standalone equipment at each stage, performing repeated Optical–Electrical–Optical (OEO) conversions, accumulating complexity and creating bottlenecks in Capital Expenditure (CAPEX) and power consumption \cite{PedroJOCN2023,Pedro:25}. To mitigate these inefficiencies, integrated IP-over-WDM (IPoWDM) architectures have emerged, leveraging coherent Point-to-Multipoint (PtMP) transmission enabled by Digital Subcarrier Multiplexing (DSCM) \cite{welch2023, Hernandez23}.

Recent studies have demonstrated DSCM-PtMP benefits in filterless environments, including Integer Linear Programming (ILP) frameworks for amplifier and coupler optimization in horseshoe-and-spur topologies \cite{Hosseini2024}, as well as traffic-driven analyses of ASIC power savings \cite{Castro2024_ECOC}. Additional work has provided statistical transceiver design guidelines \cite{Castro2025_APC}, techno-economic evaluations under traffic growth \cite{Castro2025_JOCN}, and analyses of Out-of-Band (OB) noise and Optical Amplifier (OA) optimization in filterless aggregation arcs \cite{Gatti2026}.

However, extending these benefits to large metropolitan area networks introduces new constraints. While filterless systems are primarily limited by OB noise and passive splitting losses, hierarchical deployments must also account for penalties from cascaded Wavelength-Selective Switches (WSSs) in Reconfigurable Optical Add-Drop Multiplexer (ROADM) nodes. Moreover, for longer reaches spanning metro and core segments, conventional Optical Signal-to-Noise Ratio (OSNR) metrics become insufficient for fiber nonlinear effects. This work, conducted within the EU-funded \textit{ALLEGRO} project \cite{allegro_project}, extends PtMP analysis to a large-scale reference network. The proposed architecture integrates a tree-based Access-to-Metro (AtM) aggregation segment with a ROADM-based mesh Metro-to-Core (MtC) backbone spanning up to 419~km. A Generalized Signal-to-Noise Ratio (GSNR) model based on the Gaussian Noise (GN) framework is adopted, accounting for Self-Phase Modulation (SPM), Cross-Phase Modulation (XPM), and inter-channel Stimulated Raman Scattering (ISRS). Comparing a gray-optics \textit{Benchmark} with \textit{PtP-AtM} and \textit{PtMP-AtM} IP-over-WDM (IPoWDM) scenarios, we show DSCM-enabled PtMP achieves up to 92.0\% CAPEX reduction and 99.2\% power savings compared to the traditional benchmark, offering a pathway toward 6G-ready transport.
\vspace{-0.1cm}
\section{Network Architecture and Scenarios}

\begin{figure*}
    \centering
    \includegraphics[width=0.75\linewidth]{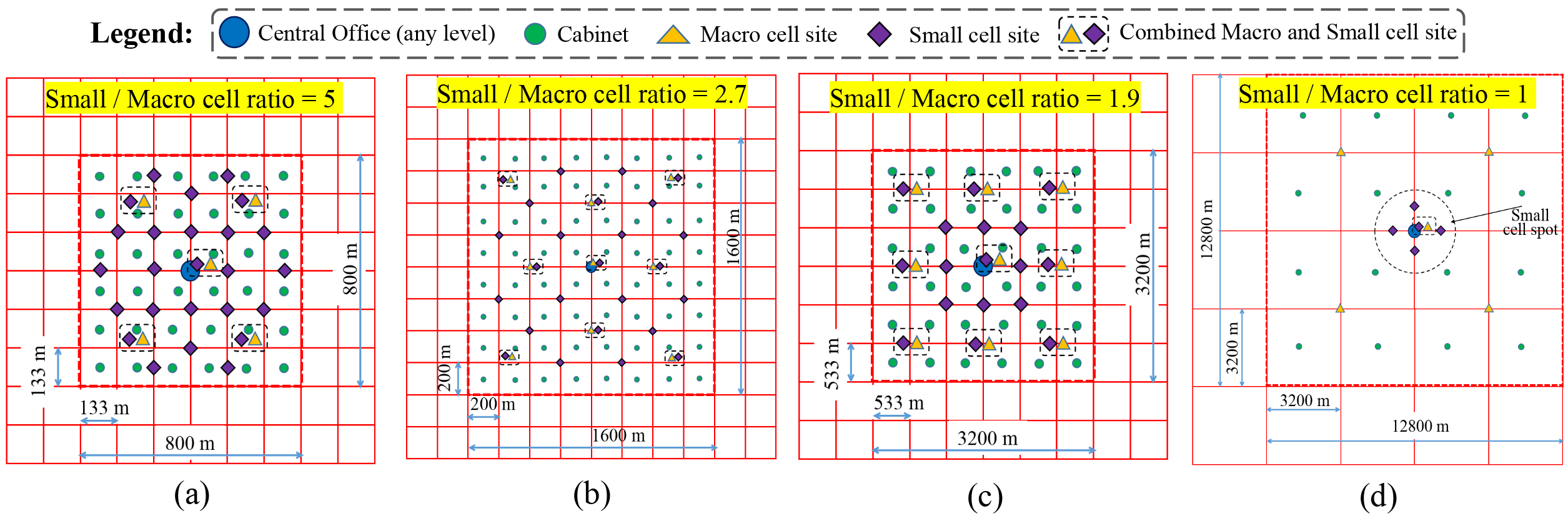}
    \caption{Network topologies for the evaluated geotypes in the ALLEGRO reference network: (a) Dense Urban, (b) Urban, (c) Suburban, and (d) Rural.}
    \label{fig:geotypes}
\end{figure*}

\begin{figure*}
    \centering
    \includegraphics[width=0.75\linewidth]{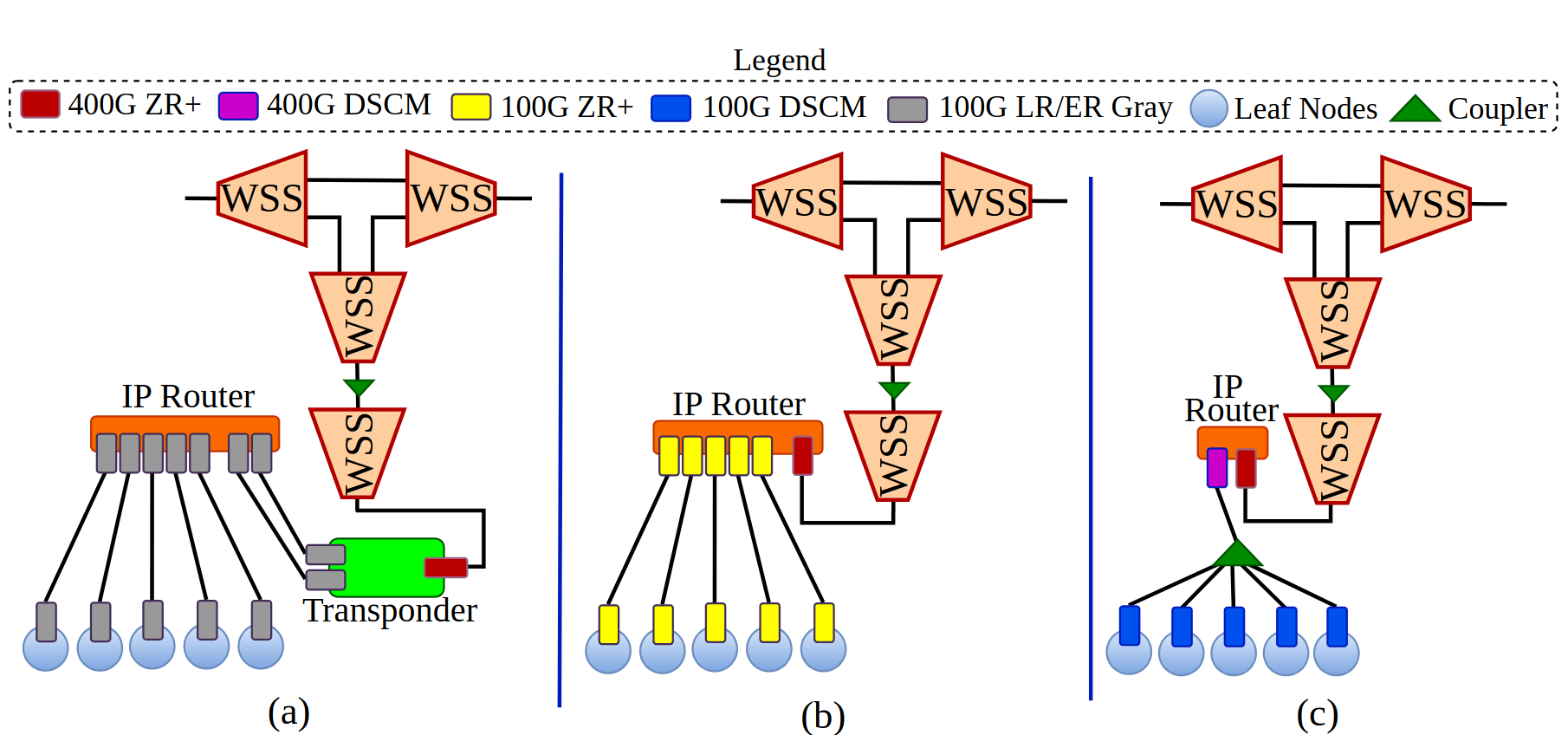}
    \caption{Node architecture of (a) Benchmark (gray) scenario, (b) PtP-AtM IPoWDM, (c) PtMP-AtM IPoWDM}
    \label{fig:node_arc}
\end{figure*}

We evaluate three architectures across AtM and MtC segments, spanning four geotypes (Fig.~\ref{fig:geotypes}). The \textit{Benchmark} (Fig.~\ref{fig:node_arc}a) uses 100G gray interfaces (LR/ER) at leaf nodes and standalone transponders at Central Offices (CO). In contrast, \textit{PtP-AtM} (Fig.~\ref{fig:node_arc}b) replaces gray optics with 100G ZR pluggables at both leaf nodes and COs, directly interfaced to IP routers, eliminating transponders; their 80~km reach \cite{photonics13020125} satisfies the 13~km maximum leaf-to-CO distance. The \textit{PtMP-AtM} architecture (Fig.~\ref{fig:node_arc}c) replaces 100G ZR with 100G DSCM transceivers at leaf nodes and uses 400G DSCM at COs to aggregate four 100G channels. For the MtC segment, where spans reach 419~km, both \textit{PtP-AtM} and \textit{PtMP-AtM} deploy 400G ZR+ modules (450~km reach \cite{photonics13020125}). All scenarios implement dual-homed protection via Link- and Node-Disjoint (LAND) paths to ensure high reliability.

\section{Quality of Transmission (QoT) Estimation}

QoT is evaluated using segment-specific metrics. For \textit{AtM}, we use 50~GHz channel spacing and 27.95~GBaud symbol rate, deploying 100G DSCM for \textit{PtMP-AtM} and 100G ZR for \textit{PtP-AtM}. Neglecting nonlinearities in short spans, performance is assessed via $OSNR = P_{rx}/P_{ASE}$, where $P_{rx} = P_{tx} - A_{total}$ and $A_{total} = \alpha L + \Gamma_{split} + A_{pol}$ ($\alpha = 0.2$~dB/km, $A_{pol} = 0.5$~dB). The splitter loss $\Gamma_{split} = 10 \log_{10}(N)$ applies only to PtMP. The Amplified Spontaneous Emission (ASE) noise is $P_{ASE} = (G-1)h\nu B F_n$, with Erbium-Doped Fiber Amplifier (EDFA) gain $G$ compensating for $A_{total}$ and $F_n = 4.5$~dB. 
In the \textit{MtC} segment, the GSNR is adopted to capture linear and nonlinear impairments. Following the GN model \cite{arpanaei2024enabling}, the GSNR accounts for chromatic dispersion and Nonlinear Interference (NLI) effects, including SPM, XPM, and ISRS, ensuring high-fidelity assessment of coherent lightpaths.

\section{Cost and Power Calculation Methodology}
Optical CAPEX and power consumption are evaluated for AtM and MtC segments, with one cost unit (c.u.) equal to 5,000~euros \cite{photonics13020125}. AtM metrics ($C_{AtM}, P_{AtM}$) are: (i)~\textit{Benchmark}: $C_{AtM}^{(1)} = 2 C_{gray}^{leaf} + 2 C_{gray}^{CO} + C_{TP}$ using 100G gray (LR: 0.08~c.u., 3.5~W; ER: 0.4~c.u., 4.5~W) and 400G Transponders (TP) (7.1~c.u., 665~W); (ii)~\textit{PtP-AtM}: $C_{AtM}^{(2)} = 2 C_{ZR}^{100G}$ utilizing 100G ZR (0.8~c.u., 5.5~W); and (iii)~\textit{PtMP-AtM}: $C_{AtM}^{(3)} = C_{DSCM}^{100G} + C_{DSCM}^{400G}$ involving 100G (1~c.u., 5.5~W) and 400G (1.2~c.u., 18~W) DSCM transceivers \cite{photonics13020125}. 

Conversely, MtC metrics are uniform: $C_{MtC} = C_{ROADM} + C_{MCS} + C_{ZR+}^{400G}$ and $P_{MtC} = P_{RoB} + P_{ZR+}^{400G}$. this segment includes ROADM-on-a-Blade (RoB) (5.7~c.u., 910~W), Multi-cast Switch (MCS) (2.1~c.u.), and 400G ZR+ (1.5~c.u., 22.5~W) \cite{arpanaei2024enabling}.

\section{Simulation Setup and Network Parameters}
The proposed scenarios are evaluated using the SEASON and ALLEGRO synthesized real-world reference networks and traffic matrices \cite{rivas2024tefnet24,allegro_project,rivas_moscoso_2025_season_access_metro}. The AtM segment consists of 876 leaf nodes aggregating traffic into 38 COs, which are interconnected via 46 fiber links (avg. 21.2~km) within the MtC segment. This architecture follows a three-layer hierarchy (2 HL3, 10 HL4, and 26 HL5 nodes) with an HL5$~\rightarrow~$HL4$~\rightarrow$~HL3 aggregation flow. Initial traffic per leaf node averages 43.6~Gbps (ranging from 10.5 to 95~Gbps) with a 40\% annual growth rate, protected via a dual-homing strategy using LAND paths. The optical layer utilizes 400G ZR+ transceivers (64~GBaud symbol rate, 75~GHz channel spacing) across a 6~THz bandwidth (C+SuperC). To ensure high-fidelity results, simulations account for a variable WSS filtering penalty (0.3--8~dB), a 1~dB aging margin, and a 36~dB back-to-back (B2B) SNR \cite{arpanaei2024enabling}.

\begin{figure*}
    \centering
    \includegraphics[width=0.8\linewidth]{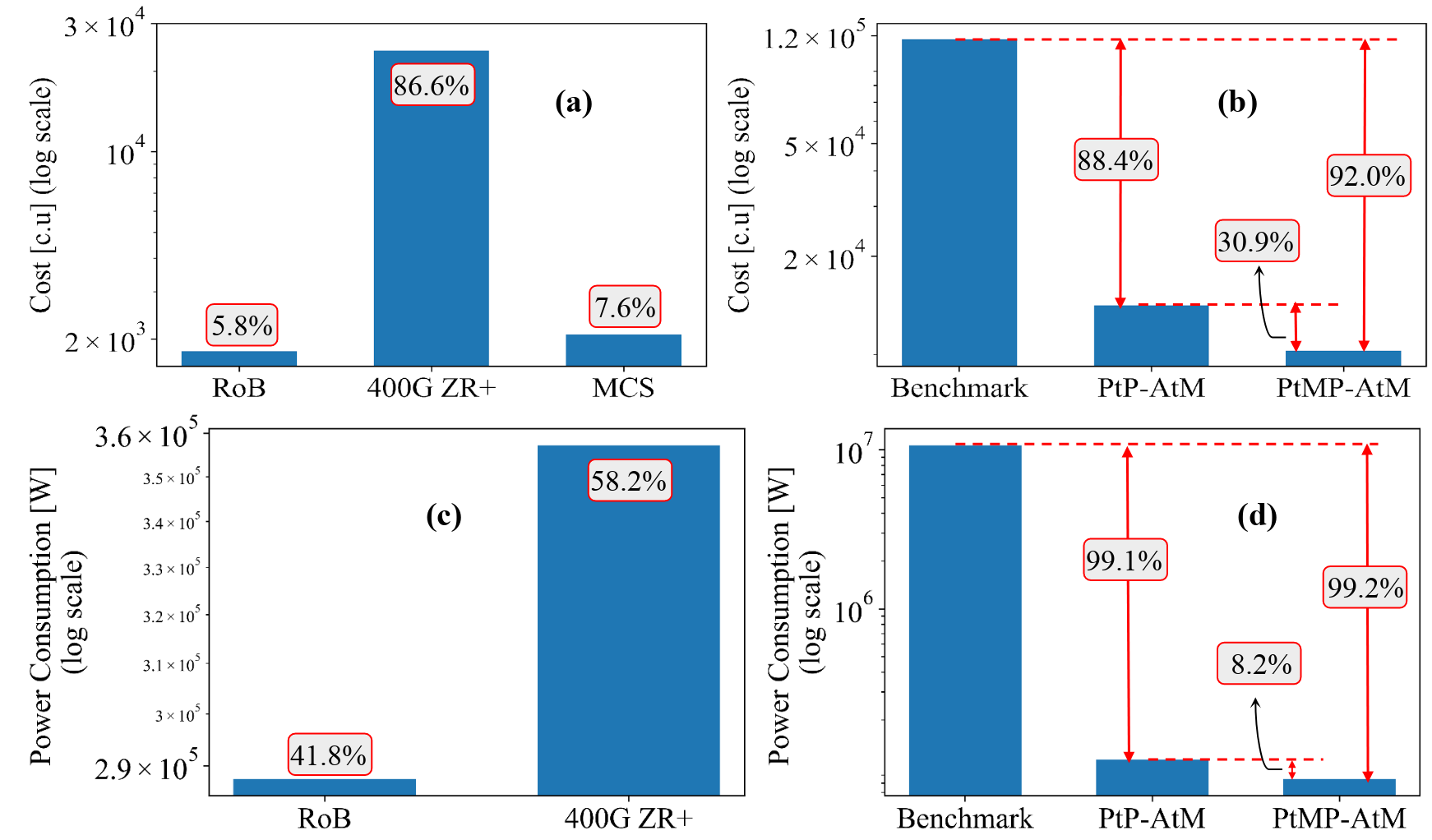}
    \caption{(a) Total cost and (c) power consumption breakdown by element in the MtC network, where percentages indicate the relative contribution of each component to the segment total. (b) Total cost and (d) power consumption in the AtM segment; here, percentages denote the relative difference between the lower and higher values across the evaluated scenarios.}
    \label{fig:cost_power}
\end{figure*}

\begin{figure}
    \centering
    \includegraphics[width=0.9\linewidth]{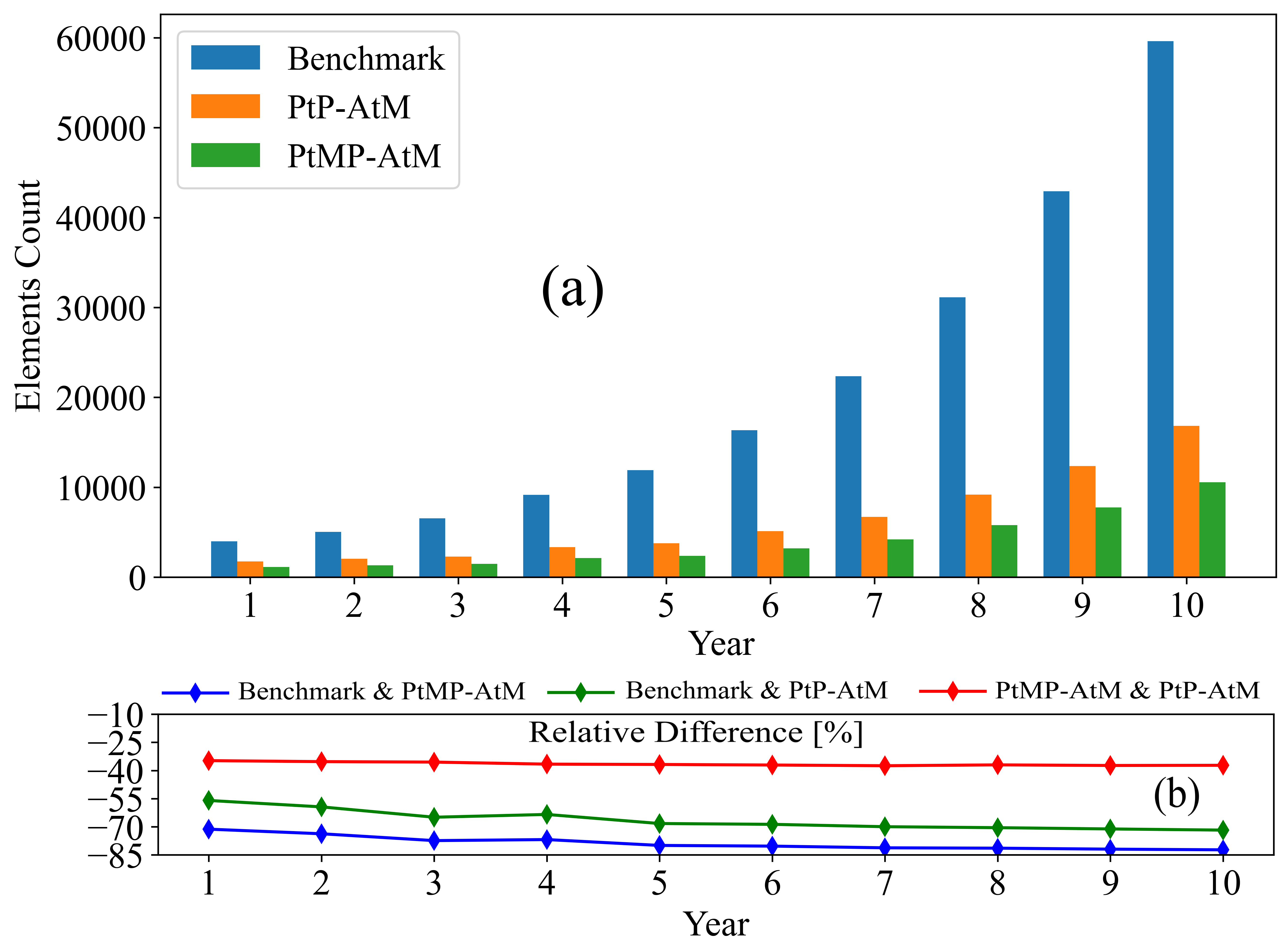}
    \caption{(a) Number of elements deployed in the AtM network under three scenarios, (b) The relative difference between scenarios.}
    \label{fig:elements-count}
\end{figure}

\section{Simulation Results}

The deployment of AtM network elements is evaluated over a ten-year horizon to compare architectural efficiency. Fig.~\ref{fig:elements-count}(a) shows the cumulative element count, while Fig.~\ref{fig:elements-count}(b) reports relative differences. In the \textit{Benchmark} scenario, 876 leaf nodes use 100G gray transceivers (860 LR, 16 ER), requiring 1,356 additional 100G LR modules and 872 CO transponders for aggregation. The \textit{PtP-AtM} replaces this hierarchy with 100G ZR transceivers at both ends, whereas \textit{PtMP-AtM} further reduces the CO footprint using 400G DSCM modules to aggregate multiple 100G DSCM leaf units. The \textit{Benchmark} remains the most hardware-intensive, requiring 71.3\% and 56.0\% more elements than \textit{PtMP-AtM} and \textit{PtP-AtM} in Year~1, increasing to 82.3\% more than PtMP by Year~10 under 40\% annual traffic growth. As shown in Fig.~2(b), \textit{PtMP-AtM} is the most efficient, using 34.8\% fewer elements than \textit{PtP-AtM} in Year~1 and 37.2\% fewer by Year~10, confirming the scalability benefits of IPoWDM, especially in PtMP form.
The economic and power consumption performance of the three scenarios are evaluated at the end of the ten-year period, with Fig.~\ref{fig:cost_power} illustrating these metrics on a logarithmic scale. In the \textit{MtC} segment, all scenarios are identical, with total cost of 27,508.4~c.u., dominated by 400G ZR+ (86.6\%), followed by MCS (7.6\%) and RoB (5.8\%), and total power of 613.9~kW, mainly from 400G ZR+ (58.2\%) and RoB (41.8\%).
In contrast, the \textit{AtM} segment shows substantial differences. The \textit{Benchmark} scenario is the least efficient, requiring 116,325.5~c.u. and 10,712.3~MW; transponder costs are included here for fair comparison with IPoWDM scenarios, where this functionality is integrated into the optical interfaces. By adopting IPoWDM, the \textit{PtP-AtM} architecture reduces cost and power consumption by 88.4\% and 99.1\%, respectively. The most optimized performance is achieved by the \textit{PtMP-AtM} architecture, which yields the lowest cost (9,306.8~c.u.) and power consumption (84.9~kW). This corresponds to a 92.0\% cost reduction and 99.2\% power saving relative to the \textit{Benchmark}, while also outperforming \textit{PtP-AtM} by 30.9\% in cost and 8.2\% in power efficiency. These results confirm that the DSCM-based Point-to-Multipoint approach is the most cost-effective and energy-efficient solution for the AtM segment.

\section{Conclusion and Future Work}

This study addresses the inherent hardware complexity and energy inefficiencies of traditional transponder-based architectures through a multi-tier network redesign. A ten-year evaluation demonstrates that the PtMP-AtM architecture achieves a 92.0\% reduction in CAPEX and 99.2\% in power consumption relative to the traditional benchmark. Furthermore, PtMP designs outperform Point-to-Point (PtP) alternatives by reducing total element counts by 37.2\% and costs by 30.9\% by the end of the horizon. These results validate DSCM-based PtMP IPoWDM as a highly scalable and sustainable paradigm for 6G-ready transport.


\section{Acknowledgements}
 The authors of UC3M and Telefonica would like to acknowledge the support of
the EU-funded ALLEGRO project (grant No. 101092766). Moreover, the UC3M authors would like to acknowledge
the support of the Spanish-funded TUCAN6-CM
project (Grant No. TEC-2024/COM-460), funded
by the Community of Madrid (ORDER 5696/2024)
and the ANNA project funded by the Spanish AEI.

\printbibliography

\end{document}